\documentclass[fleqn,12pt]{wlscirep}
\usepackage[utf8]{inputenc}
\usepackage[T1]{fontenc}
\usepackage{eucal}
\usepackage{multirow}
\usepackage{tabularx}
\usepackage{caption}
\usepackage{rotating}
\usepackage{color}
\usepackage{float}
\usepackage{lmodern}
\usepackage{bm}
\usepackage{amssymb}
\usepackage{ulem}
\usepackage{adjustbox}

\newcommand{\Bx}{\bm{x}}
\newcommand{\Ba}{\bm{a}}

\newcommand{\braopket}[3]{\langle #1 | #2 | #3\rangle}

\usepackage{wasysym}

\usepackage{wasysym}
\usepackage{amsthm}
\usepackage{subfigure}

\usepackage{soul}

\newcommand{\ket}[1]{\lvert#1\rangle} % Ket

\title{Artificial-Intelligence-Driven Shot Reduction in Quantum Measurement}

\author[1,*]{Senwei Liang}
\author[2,*]{Linghua Zhu}
\author[2]{Xiaolin Liu}
\author[1,$^\dag$]{Chao Yang}
\author[2,$^\dag$]{Xiaosong Li}
\affil[1]{Lawrence Berkeley National Laboratory, Berkeley, CA, 94720, U.S.A}
\affil[2]{Department of Chemistry, University of Washington, Seattle, WA 98195, U.S.A}
\affil[*]{these authors contributes equally to this work and are listed with alphabetical order}
\affil[$^\dag$]{Correspondence should be addressed to: cyang@lbl.gov and xsli@uw.edu}

\begin{abstract}

Variational Quantum Eigensolver (VQE) provides a powerful solution for approximating molecular ground state energies by combining quantum circuits and classical computers. However, estimating probabilistic outcomes on quantum hardware requires repeated measurements (shots), incurring significant costs as accuracy increases. Optimizing shot allocation is thus critical for improving the efficiency of VQE. Current strategies rely heavily on hand-crafted heuristics requiring extensive expert knowledge. This paper proposes a reinforcement learning (RL) based approach that automatically learns shot assignment policies to minimize total measurement shots while achieving convergence to the minimum of the energy expectation in VQE. The RL agent assigns measurement shots across VQE optimization iterations based on the progress of the optimization. This approach reduces VQE's dependence on static heuristics and human expertise. When the RL-enabled VQE is applied to a small molecule, a shot reduction policy is learned. The policy demonstrates transferability across systems and compatibility with other wavefunction ansatzes. In addition to these specific findings, this work highlights the potential of RL for automatically discovering efficient and scalable quantum optimization strategies.

\end{abstract}
\begin{document}

\flushbottom
\maketitle
\thispagestyle{empty}

\section{Introduction:}

In recent years, the rapid growth of Artificial Intelligence (AI) has brought transformative changes across various domains, including healthcare~\cite{esteva2017dermatologist,zhavoronkov2019deep}, finance~\cite{belanche2019artificial}, environmental science~\cite{rolnick2022tackling}, astronomy~\cite{fluke2020surveying} and now, the rapidly evolving field of quantum computing~\cite{perdomo2018opportunities, ciliberto2018quantum, yao2021reinforcement, krenn2023artificial}. Notable milestones in AI, such as AlphaGo's victory over world champion Go players~\cite{silver2016mastering}, the widespread use of advanced conversational agents like ChatGPT~\cite{ray2023chatgpt}, and the development of large language models~\cite{zhao2023survey,zhong2023let}, have highlighted AI's remarkable capabilities. 
These breakthroughs showcase AI's ability to not only handle and analyze vast amounts of data but also to learn from complex patterns and make decisions with a level of sophistication that was previously unattainable. This is particularly crucial in complex systems where traditional computational methods are limited due to their lack of processing power or the intricate nature of the problems~\cite{biamonte2017quantum}.

\begin{figure*}[t!]
    \centering
    \includegraphics[width=0.8\linewidth]{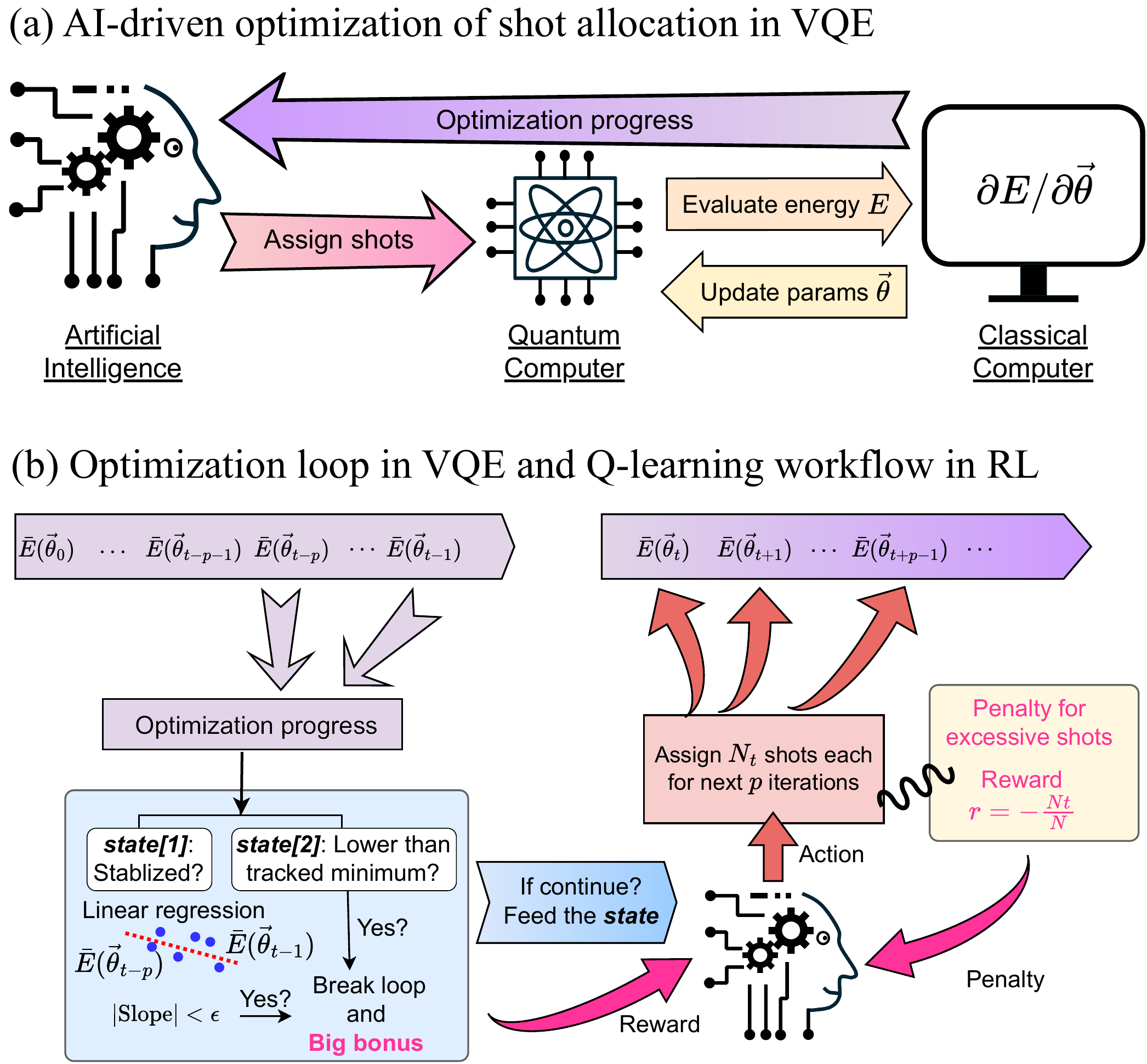}
    \caption{(a) A semantic illustration of the AI-driven optimization of shot allocation in VQE. (b) Optimization loop in VQE and Q-learning workflow in RL. }
    \label{fig:1}
\end{figure*}

Quantum computing represents a transformative leap beyond classical computing, tapping into the profound intricacies of quantum mechanics to manipulate and process data. By exploiting phenomena like superposition and entanglement, quantum computers hold the potential to tackle challenges that have long eluded classical systems. Among the myriad of applications poised to benefit from this quantum revolution, quantum chemistry stands at the forefront, promising unprecedented insights into molecular interactions and behaviors~\cite{lanyon2010towards,cao2019quantum,mcardle2020quantum}. Traditional computational methods in quantum chemistry, such as Hartree-Fock~\cite{hartree1928wave} and post-Hartree-Fock methods~\cite{bartlett1994applications,townsend2019post,shikano2021post}, can be computationally demanding, especially for systems with a large number of electrons. The computational cost associated with these methods scales poorly as the size of the molecular system increases, leading to an exponential growth in the resources required for accurate calculations~\cite{helgaker2013molecular}. 
The advent of quantum computing offers a paradigm shift in how we approach these complex quantum chemistry problems. Quantum computers operate on the principles of quantum mechanics and are naturally suited to simulate quantum systems. The Variational Quantum Eigensolver (VQE) is a hybrid quantum-classical algorithm that has been specifically designed to tackle the problem of finding the ground state energy of a molecule, which is a central problem in quantum chemistry. VQE operates by parameterizing a quantum state on a quantum processor and leveraging a classical optimizer to minimize the energy expectation value of this state in relation to a predefined molecular Hamiltonian~\cite{peruzzo2014variational, kandala2017hardware, parrish2019quantum, tilly2022variational}. By employing a parameterized quantum circuit, the VQE iteratively refines its approximation to the ground state, with classical optimization techniques guiding these refinements. This hybrid approach allows VQE to be error-tolerant and scalable, making it suitable for near-term quantum devices known as Noisy Intermediate-Scale Quantum (NISQ) computers~\cite{ preskill2018quantum ,cerezo2021variational }.

Quantum computations inherently produce probabilistic outcomes. Estimating these outcomes needs repeated measurements. One measurement is termed as ``one shot''~\cite{holevo2011probabilistic}. The precision of results is directly influenced by the number of shots. Increasing the shot count enhances result fidelity but amplifies cost of resource and processing time~\cite{pashayan2015estimating, barron2020measurement}. Given the constraints of current quantum hardware, optimizing shot allocation is crucial. Effective shot assignment not only ensures computational fidelity but also manages resource expenditure, making quantum algorithms like the VQE more practical and efficient~\cite{zhu2023optimizing,phalak2023shot,yen2023deterministic,crawford2021efficient}.

Some studies have explored shot assignment strategies for VQE that balance the measurement cost against the accuracy of approximating the ground state energy. 
To decrease the variance in estimated energy, several methods utilize term amplitudes as indicators for the variance of measurements\cite{gu2021adaptive,wecker2015progress,arrasmith2020operator}. Alternatively, some techniques manage variance thresholds to reduce the number of shots needed\cite{zhu2023optimizing}. Meanwhile, optimization-based strategies have been developed aiming at dynamically allocating shots throughout the optimization iterations, including methods that ensure real-time efficiency and are latency-awareness~\cite{ito2023latency} and robust amplitude analysis for more accurate energy estimations.\cite{johnson2022reducing} In addition, strategies that decrease inter-processor communication and enhance trainability for more effective convergence have been considered~\cite{qian2022shuffle}. Furthermore, studies on the effects of measurement shot noise suggest that intelligent parameter initialization can lessen the reliance on excessive shots, thereby enhancing optimization efficiency~\cite{scriva2023challenges}. 

Despite promising progress in VQE, current strategies heavily rely on hand-crafted heuristics that require extensive expert knowledge to design. This paper introduces an AI approach that uses reinforcement learning (RL) to automatically learn the optimal shot assignment strategy based solely on the optimization progress of VQE. RL is a powerful machine learning paradigm that learns decision-making policies through interactions with an environment, where an agent aims to maximize the cumulative rewards.  
It has shown immense potential in quantum computing for optimizing quantum circuits~\cite{yao2021reinforcement,ostaszewski2021reinforcement,fosel2021quantum}, error-robust gate-set design~\cite{baum2021experimental}, and more.

In the proposed AI-driven approach, the objective is to reduce the total shot count required for quantum measurements while ensuring consistent convergence in VQE process (see Figure \ref{fig:1} for an illustration). For each calculation of a quantum state's energy expectation value, the AI agent employs a neural network to determine the allocation of measurement shots. RL is employed to monitor progress, make an informed decision, and adaptively refine the shot allocation strategy. The adoption of RL is motivated by two key advantages. Firstly, it reduces the dependence on pre-set heuristics, facilitating dynamic and data-informed decision-making, thereby reducing the necessity for extensive domain expertise. Secondly, RL accelerates the identification and improvement of policies, potentially leading to novel and scalable strategies for handling more intricate quantum systems.

\section{Methods}\label{sec:method}

\subsection{Variational Quantum Eigensolver (VQE)}
The goal of VQE is to optimize the wavefunction of a quantum state, often the ground state of a quantum system, using the variational principle. The process for identifying the ground state energy is depicted in Figure~\ref{fig:1}a, which involves a loop of parameter optimization using a classical computer and the evaluation of an energy objective function using a quantum computer.

Consider $U(\vec\theta)$ as a circuit parameterized by $\vec\theta$. The parameterized wavefunction, used to measure energy expectation values on a quantum processor, can be represented as: $\ket{\psi(\vec\theta)} = U(\vec\theta) \ket{\psi_{\text{ref}}}$. Here, $\ket{\psi_{\text{ref}}}$ denotes the reference state,  often chosen as the Hartree-Fock solution. With the parameterized trial wavefunction defined, the energy expectation value can be computed:
\begin{equation}
  \begin{aligned}
    E(\vec\theta) = \braopket{\psi(\vec\theta)}{\hat{H}}{\psi(\vec\theta)}
    = \braopket{\psi_{\text{ref}}}{U^{\dagger}(\vec\theta) \hat{H} U(\vec\theta)}{\psi_{\text{ref}}} 
  \end{aligned}    
\label{eq:expected_energy}
\end{equation}
The ground state energy of the system can be found by minimizing the energy expectation $\text{min}_{\vec\theta} E(\vec\theta)$.

By evaluating the energy objective function $E(\vec\theta)$ on a quantum computer, the optimization can be performed on a classical computer. Various optimizers, such as gradient descent, Adam~\cite{Kingma2014AdamAM} and the Broyden-Fletcher-Goldfarb-Shanno method~\cite{fletcher2013practical} can be utilized to iteratively update and refine the parameter $\vec\theta$.

The energy objective function in Eq.~\ref{eq:expected_energy} is evaluated on a quantum computer by taking multiple measurement shots.
Each measurement (shot) of $E(\vec\theta)$ can be viewed as taking a sample, $e^s(\theta)$, of a random variable with the mean $E(\vec\theta)$ and a standard deviation $\sigma(\vec\theta)$. By taking multiple samples, $e^s(\theta), s=1, 2, ..., N$, we have $\bar{E}(\vec\theta):=\frac{1}{N}\sum_{s=1}^Ne^s(\theta)\approx E(\vec\theta)$. %If we perform multiple measurements to obtain $e^s(\theta)$, $s=1, 2, ..., N$ as the energy estimations of $E(\theta)$, $E(\theta)$ can be estimated  by taking the average of $e^s(\theta)$'s to yield the empirical mean 
%\begin{equation}
%\hat{E}(\theta) = \frac{1}{N} \sum_{s=1}^{N}e^s(\theta)\approx %E(\theta).
%\label{eq:emean}
%\end{equation}
Based on the central limit theorem~\cite{billingsley2017probability}, the probability of $\bar{E}(\vec\theta)$ deviating from $E(\vec\theta)$ decreases as the number of measurements $N$ increases. However, increasing $N$ also incurs a larger resource cost.

When measuring the energy expectation value, two primary factors need to be taken into account: (1) the total number of measurement shots $N$ and (2) the assignment of these shots among the individual terms or cliques of the Hamiltonian.

To address the first primary factor, we aim to develop an RL-driven policy capable of making informed decisions about the total number of measurement shots $N$ (see the next section). The RL policy will be compared with the Variance-Preserved Shot Reduction (VPSR) strategy, which aims to minimize the total number of shots while maintaining a specified target variance in the measurements of the energy objective function\cite{zhu2023optimizing}.
The primary distinction between the VPSR and RL strategies lies in their shot reduction techniques: the RL strategy leverages an RL-policy for optimizing the total number of shots, whereas VPSR is grounded in a mathematical lower-bound for shot reduction, the specifics of which are elaborated in Ref. \citenum{zhu2023optimizing}.

After the total number of measurement shots $N$ is determined by an RL policy, various shot assignment strategies can be employed. To address the second primary factory, we will compare two classes of shot assignment approaches. The first method, called ``Uniform'', consistently distributes the same number of measurement shots to each Hamiltonian clique throughout all iterations. In contrast, the second method, known as Variance-Minimization (VM) shot allocation, adopts a dynamic shot allocation strategy based on minimizing variance in the measurements of the energy objective function, varying the number of shots for different Hamiltonian cliques accordingly.
It is known that the variance in the estimation of energy objective function is minimized if and only if the shot allocation ratios are proportional to the standard deviations of measurements of the terms/cliques, {\it i.e.},
\begin{align}
N_1:N_2:\cdots:N_m&=\sigma_1(\vec\theta):\sigma_2(\vec\theta):\cdots:\sigma_m(\vec\theta)
\label{eqn:ratio}
\end{align}
By allocating shots according to this ratio, one can minimize the variance of the Monte-Carlo estimate of the total energy given a fixed shot budget. Consequently, the selection of the total number of measurement shots becomes a critical decision, impacting the overall efficiency and efficacy of the quantum computing task.

\subsection{Reinforcement Learning of Measurement Policy}

The objective of this research is to create an AI agent that employs an optimal policy for making informed decisions about the total number of shots during the VQE optimization process. To achieve this, we implement a reinforcement learning (RL) approach, which is designed to learn the optimal shot selection policy through iterative training episodes. In the context of quantum computing measurements, each VQE execution represents a training episode for the neural network that forms the core of the RL-policy, as illustrated in Figure~\ref{fig:1}.

\subsubsection{Foundation of the Reinforcement Learning}
During the RL training, a state, denoted as $\Bx$, is produced in a VQE step.
The RL agent takes an action $\Ba$ in the current state $\Bx$ and a reward $r(\Bx, \Ba)$ is evaluated. The VQE process then transitions to a new state $\Bx'$. This continues until the end of the episode and yields a set of quadruples $(\Bx, \Ba, r, \Bx')$ representing the full RL and VQE interaction experience. 

The policy which the RL agent follows to take an action at a particular state $\Bx$ is designed to maximize the expectation of a sequence of discounted future rewards,
\begin{equation}
\mathbb{E}_{\tau\sim\pi} \left[\sum_{t=1}^\infty \eta^t r(\Bx^t, \Ba^t) | \Bx^0=\Bx, \Ba^0 = \Ba \right],
\end{equation}
where $\tau$ denotes an interaction sequence $\tau:=\{(\Bx^0, \Ba^0),\cdots, (\Bx^t, \Ba^t), \cdots\}$ initialized at $\Bx^0=\Bx, \Ba^0=\Ba$ and produced from the policy $\pi$. A new state $\Bx^t$ will be prepared after taking an action $\Ba^{t-1}$, which is based on the previous state $\Bx^{t-1}$. Here, $0 < \eta \leq 1$ is a discount factor. This expected return is referred to as the $Q$-function, denoted $Q^{\pi}(\Bx, \Ba)$, which represents the discounted future rewards for taking action $\Ba$ in state $\Bx$ under policy $\pi$. The action function $A(\Bx)$ gives the optimal action that maximizes the $Q$-function for the observed state $\Bx$ under policy $\pi$; formally, $A(\Bx) = \arg\max_{\Ba} Q^{\pi}(\Bx, \Ba)$.~\cite{sutton2018reinforcement}

%The optimal $Q$-function $Q^*(\Bx, \Ba):=\max_{\pi}Q^{\pi}(\Bx,\Ba)$ satisfies the Bellman equation~\cite{sutton2018reinforcement}
%\begin{equation}
%Q^*(\Bx, \Ba)={\mathbb{E}}_{\Bx^{\prime}}\left[r(\Bx, \Ba)+\eta \max _{\Ba^{\prime}} Q^*\left(\Bx^{\prime}, \Ba^{\prime}\right)\right],
%\label{eqn:bellman}
%\end{equation}
%where the expectation is taken with respect to the conditional probability of the new state $\Bx^{\prime}$ given $\Bx$ and $\Ba$.  Associated with this optimal $Q$-function is the optimal action
%\begin{align}
%    A^*(\Bx)=\arg \max _{\Ba} Q^*(\Bx, \Ba).
%    \label{eqn:optimalaction}
%\end{align}

Finding the optimal $Q$-function is often an intractable problem, and approximate solutions are commonly used instead. Several methods like Deep Deterministic Policy Gradient (DDPG~\cite{silver2014deterministic}) and Twin Delayed DDPG (TD3~\cite{fujimoto2018addressing}) have been developed to approximate the solution. In these methods, the $Q$-function $Q(\Bx,\Ba)$ and action function $A(\Bx)$ are represented by neural networks with parameter sets $\Psi$ and $\Phi$ respectively (where $\Psi$ and $\Phi$ denote all trainable weights and biases). These neural networks are trained on a dataset $\mathbb{P}$ containing tuples of the form $(\Bx, \Ba, r(\Bx, \Ba), \Bx')$. The training of parameters $\Psi$ and $\Phi$ 
%are optimized in an alternating fashion by 
%objectives derived from Eq. \ref{eqn:optimalaction} and the Bellman equation (Eq. \ref{eqn:bellman}). Specifically, the training 
solves the following optimization problems alternately:
\begin{equation}
    \max_{\Phi}\underset{\left(\Bx, \Ba, r, \Bx^{\prime}\right) \sim \mathbb{P}}{\mathbb{E}} Q(\Bx, A(\Bx;\Phi); \Psi)
    \label{eqn:qobj1}
\end{equation}
and 
\begin{equation}
    \min_{\Psi}\underset{\left(\Bx, \Ba, r, \Bx^{\prime}\right) \sim \mathbb{P}}{\mathbb{E}}\left[Q(\Bx, \Ba; \Psi)-\left(r(\Bx,\Ba)+\eta Q\left(\Bx^{\prime}, A(\Bx^{\prime};\Phi); \Psi\right)\right)\right]^2.
    \label{eqn:qobj2}
\end{equation}
Once the parameters $\Psi$ and $\Phi$ are optimized through training, the learned action function $A(\Bx;\Phi)$ can be used as the policy to interact with the VQE process.

\subsubsection{Implementation Details}
In this section, we detail the implementation specifics within the framework of the VQE optimization process (as shown in Figure~\ref{fig:1}b). Our focus is on applying the RL approach effectively to learn the optimal policy for shot selection. 

Let $N$ and $N_t$ respectively represent the maximum (the budget) and AI-selected number of shots for each energy evaluation. The goal of the RL training is to learn an optimal policy that minimizes $\sum_{t}N_t$ as the VQE process converges/optimizes the energy objective function $\bar{E}(\vec\theta)$.

{\bf State:} We use the state $\Bx$ to track the progress of the VQE optimization process. During the early stages of optimization, the algorithm is expected to explore the energy landscape in search of lower energy values. In the later stages, the energy objective function converges to a minimum, and we expect the energy values to stabilize or vibrate around the minimum. Therefore, we can characterize the optimization process by two key metrics that evaluate the progression towards the minimum: (1) the level of convergence in the optimization, and (2) the achievement of a lower minimum energy value. During the $t$-th iteration, we consider a series of energy values $\{\bar{E}(\vec\theta_{i})\}_{i=t-p}^{t-1}$. The first step involves conducting linear regression on these values against their respective iteration indices. 
A small absolute value of the regression slope indicates that the energy values have stabilized or are vibrating around a constant value, while a large absolute value of the slope indicates an ongoing trend of increasing or decreasing in the energy objective function.
For the second metric, we monitor the lowest estimated energy value from prior iterations. If the current estimated energy undercuts this minimum, it signifies a successful attainment of a lower energy value. Therefore, the state $\Bx$ is composed of these elements to gauge the optimization's effectiveness.:
\begin{align}
    \Bx\text{[1]} = -\log_{10}(|\text{slope}|), \quad \quad \Bx\text{[2]} = \left\{ \begin{aligned}1,& \quad \text{if the current} \leq \text{the tracked minimum,} \\ 0,& \quad \text{if the current} > \text{the tracked minimum.}\end{aligned}\right.\label{eq:state}
\end{align}
%These two components in the state are used to characterize the progress of optimization.

{\bf Action:} In each VQE iteration, the RL agent's responsibility is to determine the number of shots $N_t(\Bx)$ based on the state $\Bx$. Since $N_t(\Bx)$ should not exceed the maximum number of shots for each energy evaluation, the ratio $\frac{N_t}{N} \in (0, 1]$. Therefore, we define the action function $N_t(\Bx) := N \cdot A(\Bx)$, where $A(\Bx) \in (0, 1]$ represents the portion of $N$ to be used. Once $N_t(\Bx)$ is determined, it will be applied to the next $p$ iterations, with $N_t(\Bx)$ shots for each energy evaluation. Then, a new state will be calculated based on the energy values obtained from those $p$ iterations.

{\bf Reward/Penalty and Stopping Rule:} Since our goal is to learn a policy that minimizes $\sum_{t}N_{t}$ while ensuring the convergence of the energy objective function, the idea behind designing the reward in the RL policy training is to penalize the number of shots used in VQE and reward convergence of the energy objective function. The reward for taking action $N_t(\Bx)$ is defined as $r:=-\frac{N_t(\Bx)}{N}=-A(\Bx)$, which falls within the range of $[-1, 0)$. This means more penalty is applied when the agent uses a larger portion out of the budget ($N$ shots). In addition, if the convergence criteria are met, the VQE run will end and a large bonus is assigned to the agent so that the overall accumulated reward is positive. A VQE run is considered converged when $\Bx\text{[1]}$ in Eq. \ref{eq:state} is larger than a predefined threshold specific to the problem, chosen to be $3.5$ in this work, and $\Bx\text{[2]}=1$. 
When both criteria are fulfilled, we terminate the VQE optimization loop, and the agent is rewarded with a large bonus of 20 for guiding the VQE towards the variational minimum.

%\subsection{AI-Driven Optimization of Shot Assignment in VQE}

%The AI agent assigns shots for every iteration based on the learned action function $A(\Bx;\Phi)$. 

\section{Results}\label{sec:result}

In the AI-guided VQE methods, an RL policy, implemented through a neural network, is employed to dynamically determine the total number of measurement shots. This enhancement is applied to both the Uniform and VM approaches, resulting in two distinct strategies: RL-Uniform and RL-VM. While RL-Uniform retains the consistent shot distribution across all Hamiltonian cliques characteristic of the original Uniform method, RL-VM adopts a variance-minimization approach for shot allocation, similar to the standard VM. 
To demonstrate the efficacy of RL in learning the optimal strategy for the AI-driven shot reduction  approach, we carried out VQE optimizations of four molecular wavefunctions.

\subsection{Reinforcement-Learning Shot Reduction Policy using H$_2$} 

The shot allocation policy is RL-trained using VQE optimization of an H$_2$ ground state wavefunction.  A 2-parameter hardware-efficient ansatz~\cite{choy2023molecular} is used for H$_2$, mapped on to 2 qubits. Each qubit represents a spin orbital. The agent is trained for 200 episodes using the TD3 method~\cite{fujimoto2018addressing}. The agent allocated shots every 10 iterations during the VQE optimization. The uniform shot assignment strategy is used in the RL-training process.

\begin{figure*}[t!]
    \centering
    \includegraphics[width=0.6\linewidth]{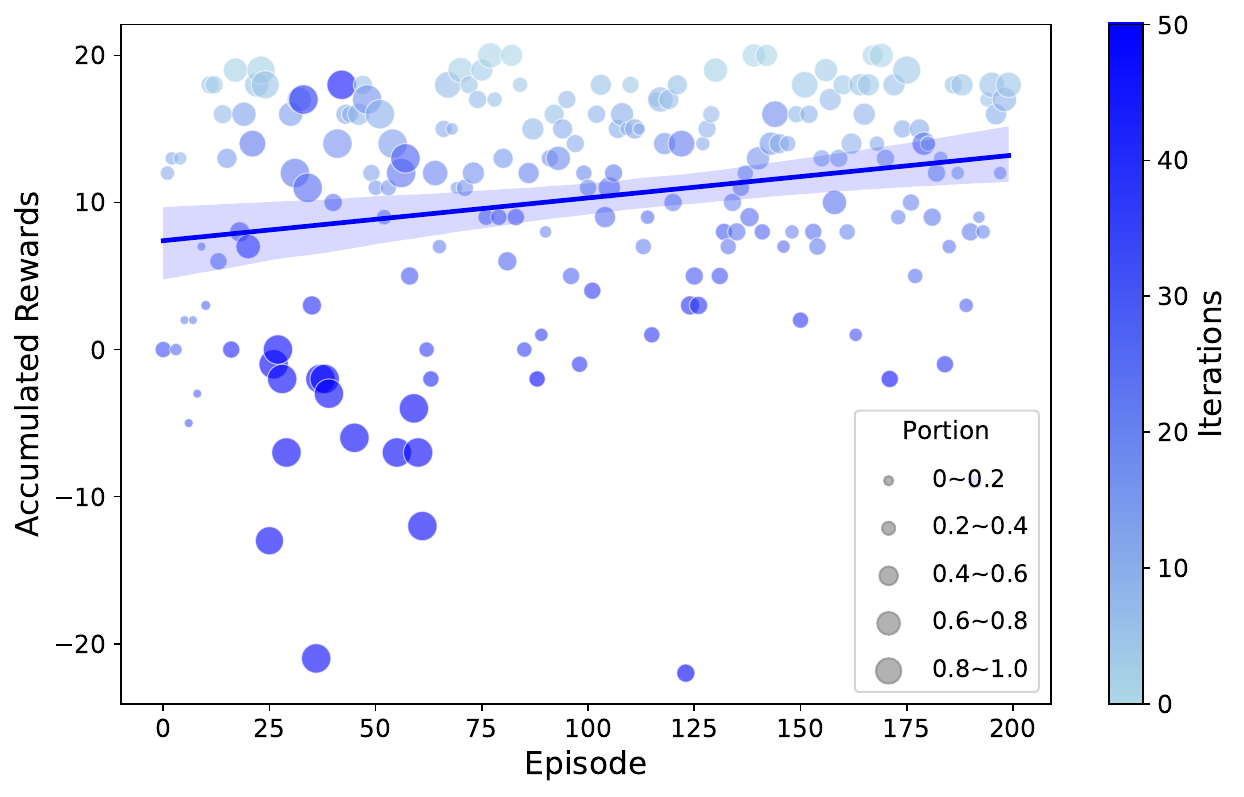}
    \caption{Scatter plot of the total rewards versus the episode number in the RL process. A linear regression line illustrates the overall trend of increasing rewards as the agent gains more experience with each episode. The circles are colored differently to indicate the number of iterations used in each episode to satisfy the termination (convergence) criteria and terminate the VQE run. The size of circle reflects the portion of shot budget used, where a circle size of 1 indicates using the entire shot budget, and a smaller size indicates that fewer shots were used.
}
    \label{fig:2}
\end{figure*}

Figure \ref{fig:2} illustrates the RL-policy training using VQE optimizations of the ground state wavefunction of H$_2$ with a bond length of $R=1.75 \AA$. 
We track how rewards increase as the agent learns over episodes. The plot includes a linear regression line to emphasize the growing trend in rewards, indicating that the agent becomes more effective with each episode in improving the policy. The color of the circle on the plot represents the number of iterations used in an episode to meet the convergence criteria. Additionally, the size of the circles indicates the most frequently utilized portion of the shot budget across iterations within a given episode of a VQE run. At the beginning of training, the agent's use of shots varies significantly—sometimes it uses only a few and other times it nearly uses the entire budget.
As training goes on, the agent starts using a more consistent amount of shots, generally between 30\% and 65\% of its total available shots. This change shows that the agent is developing a smarter, more balanced approach and getting a better grasp of how to tackle the problem effectively.

%How does the policy change? How to quantify the change?

\begin{figure*}[t!]
    \centering
    \includegraphics[width=0.8\linewidth]{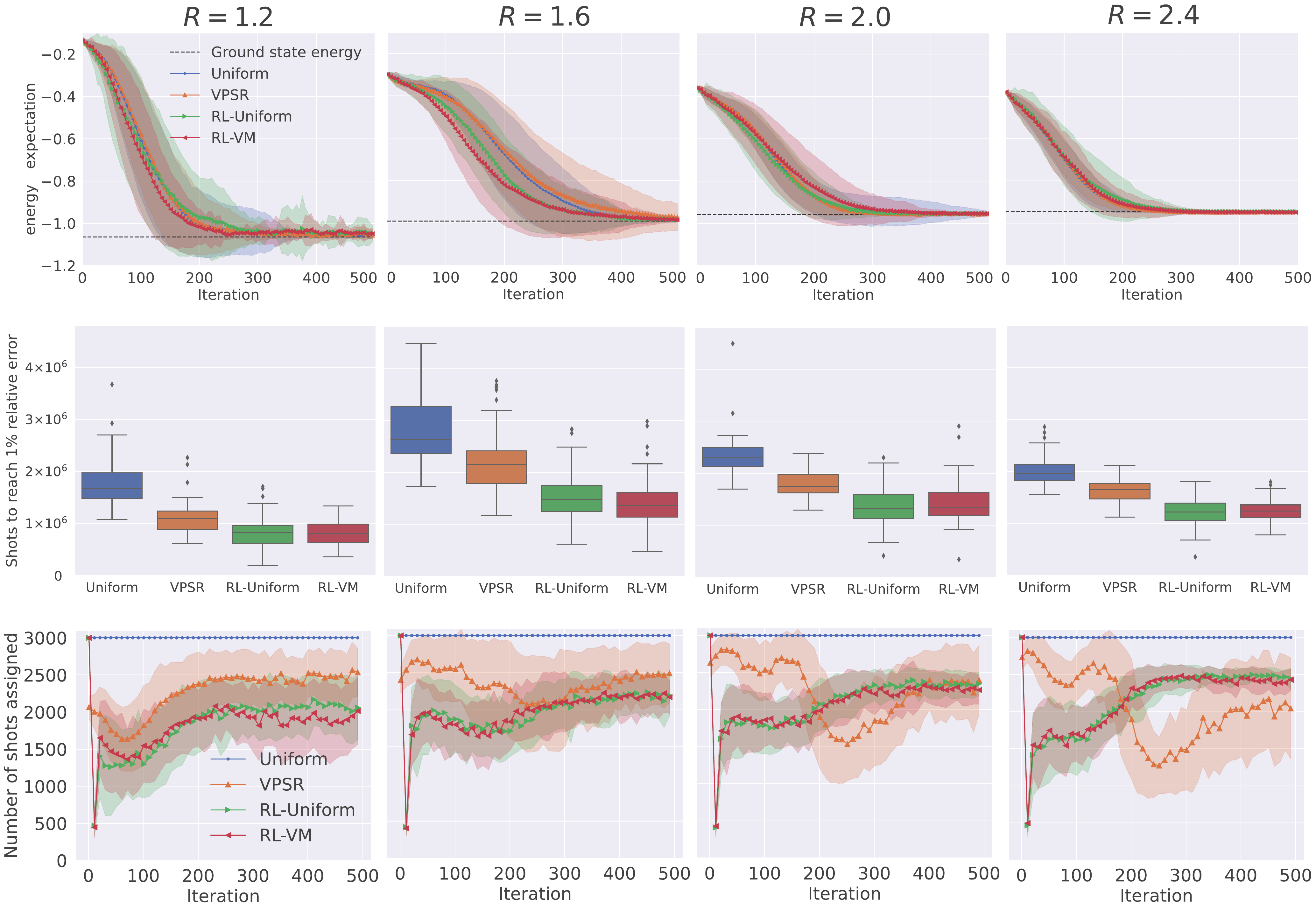}
    \caption{VQE optimizations of H$_2$ at different bond lengths. In each VQE run, the ansatz parameters is optimized using gradient descent with a learning rate of $0.1$ for $T=500$ iterations. A maximum budget of $N=3000$ shots is used for each energy evaluation. All experiments were conducted using the Qiskit QasmSimulator. Each result is based on 60 independent experiments. \textbf{Top}: the energy (Hartree) is plotted as a function of the number of iterations. \textbf{Middle}: boxplots that summarize the number of shots needed to achieve an energy with an error under 1\% of the ground state energy. These boxplots facilitate a comparison of the distributional differences between datasets. Within each boxplot, the five horizontal lines, from bottom to top, represent the minimum, 25th percentile, median, 75th percentile, and maximum values. Black dots are used to denote outlier data points. \textbf{Bottom}: the distribution of shots across each iteration.}
    \label{fig:3}
\end{figure*}

\subsection{Transferability of the RL-Learned Shot Reduction Policy} 

To evaluate the transferability of the AI-derived shot-reduction strategy, we implemented the RL-policy, initially learned from the VQE optimization of H$_2$ at a bond length of $R_{\text{H}_2}=1.75~\AA$, across various other bond lengths. The results are shown in Figure \ref{fig:3}. In these quantum computing simulations, the RL-learned policy for $R_{\text{H}_2}=1.75~\AA$ is held constant for other bond lengths. The top panel of Figure \ref{fig:3} shows that all methods considered here can successfully optimize the wavefunction of H$_2$ at different bond lengths with a similar convergence profile. The middle panel of Figure \ref{fig:3} shows that the RL-Uniform and RL-VM approaches using the transferred RL-policy from $R_{\text{H}_2}=1.75~\AA$ can substantially decrease the total number of shots required to achieve convergence at other bond lengths. The bottom panel illustrates that the total shots allocated by RL-learned policies vary throughout the VQE optimization, highlighting the adaptive nature of the shot allocation process. The RL-Uniform method exhibits an impressive capability in reducing the number of shots needed to achieve convergence. Across all bond lengths, we observe a significant saving in the shot budget, averaging between 40-55\%, compared to the conventional Uniform approach. 

 \begin{figure*}[t!]
    \centering
    \includegraphics[width=0.8\linewidth]{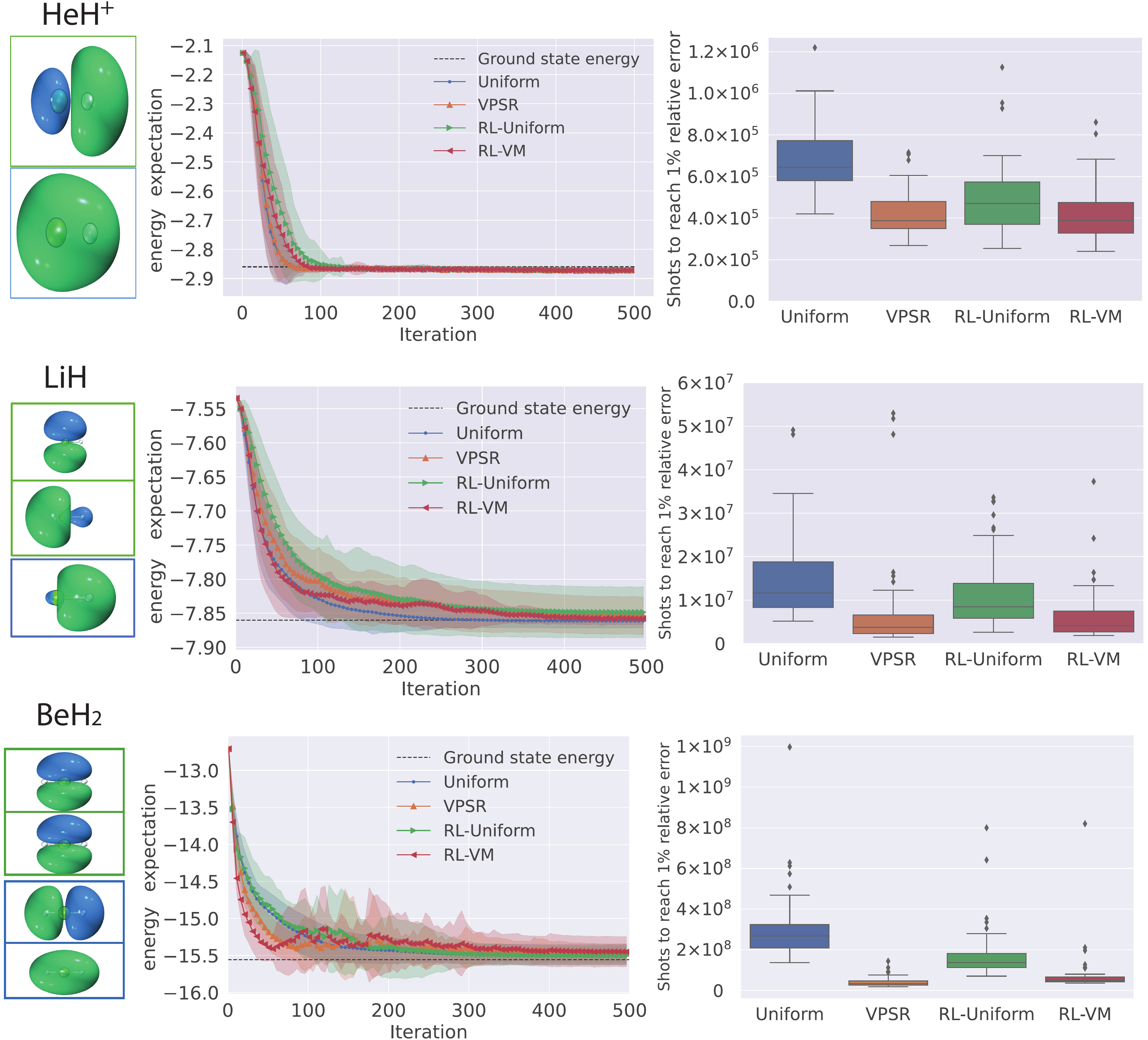}
    \caption{\text{The HeH$^+$ (top), LiH (middle) and BeH$_2$ (bottom) molecule systems.} In each VQE run, the ansatz parameters are optimized using Adam with a learning rate of $0.1$ for $T=500$ iterations and the learning rate follows a cosine decay as iteration increases. A maximum budget of $N=4000$ shots for HeH$^+$,  $N=27000$ shots for LiH and $N=96000$ shots for BeH$_2$ is used for each energy evaluation. All experiments were conducted using the Qiskit QasmSimulator. Each result is based on 60 independent experiments. \textbf{First column}: molecular orbital plots. \textbf{Second column}: the energy (Hartree) is plotted as a function of the number of iterations. \textbf{Third column}: boxplots that summarize the number of shots needed to achieve an energy with an error under 1\% of the ground state energy. These boxplots facilitate a comparison of the distributional differences between datasets. Within each boxplot, the five horizontal lines, from bottom to top, represent the minimum, 25th percentile, median, 75th percentile, and maximum values. Black dots are used to denote outlier data points.}
    \label{fig:4}
\end{figure*}

% \begin{figure}[ht]
% \centering
% \subfigure[HeH]{\includegraphics[width=0.46\linewidth ]{HeH_figures/HeH_3D_energy_vs_iters_adam.pdf}}
% 	\subfigure[HeH]{\includegraphics[width=0.46\linewidth ]{HeH_figures/HeH_3D_boxplot_adam.pdf}}
% 	\subfigure[LiH]{\includegraphics[width=0.46\linewidth ]{LiH_figures/LiH_8D_energy_adam.pdf}}
% 	\subfigure[LiH]{\includegraphics[width=0.46\linewidth ]{LiH_figures/LiH_8D_boxplot_adam.pdf}}
% \caption{New results for LiH and HeH. The only difference is that previously we used gradient descent but this plot uses Adam optimizer. \textcolor{red}{It seems that it is better than the results in Figure 4 and we could adpot this one. }}
% \end{figure}

To further evaluate the transferability and efficacy of the RL policy, we applied the RL shot-reduction strategy, initially trained on the H$_2$ molecule, to the VQE optimizations for the HeH$^+$, LiH, and BeH$_2$ molecular wavefunctions. For HeH$^+$ and LiH, the bond lengths were set at 1.00~\AA{} and 1.45~\AA, respectively. The HeH$^+$ system utilized a 3-parameter unitary coupled-cluster with singles and doubles (UCCSD) ansatz, as described in Ref. \citenum{gunther2021improving}, while the LiH employed a two-layer hardware-efficient ansatz featuring a sequence of single-qubit rotations and entangling gates, accounting for 8 variational parameters, following the approach outlined in Ref. \citenum{choy2023molecular}.

The Hamiltonian for BeH$_2$ was bulit from the STO-3G fermionic basis set, with 1s, 2s, and 2p$_x$ orbitals of beryllium, and 1s orbitals of hydrogen, totaling 10 spin orbitals. To simplify the model, the innermost 1s orbitals of beryllium were considered fully occupied.
% and were energetically adjusted. This adjustment was achieved by diagonalizing the non-interacting part of the fermionic Hamiltonian.
The system was further reduced to a six-qubit model via spin-parity mapping and qubit tapering, setting the molecular geometry with a Be-H bond length of 1.5~\AA{}~\cite{kandala2017hardware}. A `full 2-depth RyRz ansatz' with 24 parameters was implemented for the quantum simulations of BeH$_2$~\cite{tkachenko2021correlation,ferrari2021deterministic}.

This setup allowed us to thoroughly investigate the RL policy's transferability across different molecular systems and varying qubit configurations, assessing its performance on wavefunctions mapped onto 2, 4, and 6 qubits for HeH$^+$, LiH, and BeH$_2$ respectively. The results are shown in Figure \ref{fig:4}.

The second column of Figure \ref{fig:4} presents the VQE convergence profile as a function of the VQE iteration number. All methods exhibit a smooth convergence profile.
The third column of Figure \ref{fig:4} illustrates the distribution of shot counts required to achieve an estimated expectation energy within 1\% error relative to the ground state energy over 60 independent trials. In these systems, the uniform distribution method requires the most measurement shots. For HeH$^+$, RL approaches utilize the fewest measurement shots, suggesting the transferability of the RL-learned policy. 
Notably, for the two larger cases LiH and BeH$_2$, although RL methods achieve an impressive reduction of approximately 62-77\% in the total number of measurement shots compared to the uniform distribution approach, they are not always the most optimal method. The VPSR approach that is based on an analytical expression to estimate the lower-bound of the shot reduction performs  better for the BeH$_2$ test case with an 86\% shot reduction. This observation suggests that while the RL-learned policy can be transferred to a different system, it might not be the most optimal policy, or the policy may not have been trained with a sufficiently diverse dataset.

These results suggest that the AI-generated policy is proficient in optimizing the number of measurement shots required for VQE applications. Moreover, the RL-policy demonstrates a significant level of transferability, adapting effectively to various bond lengths, distinct molecular systems, different wavefunction ansatzes, and quantum systems represented by a different number of qubits. A comprehensive discussion, along with the underlying rationale for these observations, will be elaborated in the subsequent section.

%\begin{figure*}[ht]
%\centering
%\includegraphics[width=0.5\linewidth ]{figure4.pdf}
%\caption{Scatter plot of the accumulated rewards versus the episode number in the RL process. Each point represents the total reward gained during one episode. A linear regression line (solid) shows the general trend of increasing rewards as the agent gains more experience across episodes.}
%\label{fig:7}
%\end{figure*}

\section{Discussion}\label{sec:discussion}

While RL provides an automatic way to find good policies to reduce shot usage for convergence, some aspects should be considered to improve the utilization of RL, including the training expense and the robustness of RL learned policies.

\subsection{Cost of Policy Learning}

A VQE run for each RL episode requires considerable quantum measurements. This makes RL training for multiple episodes prohibitively expensive in terms of shots. In the implementation, a maximum of 200 episodes are allowed. However, as shown in Figure~\ref{fig:2}, the rewards tend to increase as the episode number increases. This suggests that the RL process continues to find better policies by extending the number of episodes. However, this can be very costly. To reduce the training cost, one can apply more data-efficient RL algorithms~\cite{lin2021continuous} that reuse experience across episodes to learn good policies from fewer full VQE runs. Furthermore, as demonstrated in this paper, the learned policy from a smaller system can be versatile, so transfer from surrogates can help find shot-economic policies. We can initialize policies for larger VQEs by first finding good allocation strategies on smaller systems and cheaper simulators. 

\subsection{Transferability of RL-Policy} 
This paper presents a single policy acquired through RL, which has been numerically shown to demonstrate transferability across various systems and compatibility with variance-minimization shot assignment strategies. The potential for transferability of the RL-policy is closely linked to the characteristics of the quantum wavefunctions examined in this study. Despite the variance in ansatzes and the mapping onto different numbers of qubits, the three test systems share a notable similarity in their ground state wavefunctions, particularly in terms of the nodal structure. Consequently, we anticipate that the policy developed here should be effectively transferable, especially for optimizing their ground state wave functions. To improve the policy for other types of molecular systems, the RL-training can be carried out with a sufficiently diverse dataset.

Although transferring the RL-policy learned on H$_2$ to the LiH and BeH$_2$ systems is effective, it does not achieve the optimal shot reduction strategy. This is mainly due to the significant difference in the density distribution. This suggests that transferring the RL-policy to quantum systems exhibiting more complex wavefunction characteristics may challenge the direct applicability of the policy. Nonetheless, the strength of RL lies in its adaptive nature, improving the policy through continuous interaction with the environment. In such scenarios, initiating the VQE process with the pre-learned RL-policy is a viable strategy. Subsequently, engaging the RL mechanism during the VQE optimization allows for iterative refinement and improvement of the policy.

\subsection{Robustness of RL-Policy} 
The RL process can yield many different policies based on the chosen RL environment design and hyperparameters (such as reward and termination criteria). These newly discovered policies should also undergo testing to evaluate their versatility and compatibility. Furthermore, it would be valuable to test them in more challenging scenarios such as quantum machine learning optimization. The current results showcase one viable policy, but future work should systematically explore the space of possible policies to identify optimal and robust strategies across diverse quantum optimization problems.

%\begin{figure*}[t!]
%    \centering
%    \includegraphics[width=\linewidth]{figure1_0123.pdf}
%    \caption{\textbf{(a)} Optimization loop in VQE and Q-learning workflow in RL.}
%    \label{fig:6}
%\end{figure*}

\section{Conclusion}\label{sec:conclusion}

This paper introduces a reinforcement learning framework designed to learn policies that minimize the number of quantum measurements (shots) needed in the VQE process to approximate the ground state energy. We developed a new RL environment that reflects the progression of VQE, where each episode corresponds to a complete run of the VQE. The environment's state is defined by the current status of the optimization. The key aspect of our approach is to balance the negative rewards for excessive shot use against significant bonuses for achieving convergence. 

From simulations with a simple H$_2$ molecule, we derived an RL policy. We further demonstrate numerically that this policy is transferable across different quantum systems, including H$_2$ at varying bond lengths, as well as HeH$^+$, LiH, and BeH$_2$ systems. It also shows compatibility with various shot assignment strategies.

Incorporating AI, particularly through reinforcement learning, into the VQE process can drastically lower the cost of essential quantum measurements. By leveraging AI to strategically reduce the number of shots, we significantly cut the overall expense and resource consumption in quantum computing experiments. This work not only makes quantum computing more accessible but also enhances its effectiveness in various applications, like material science and pharmaceuticals, by enabling more precise simulations with fewer resources.

%\section*{Some appendices}

% \begin{figure*}[ht]
% \centering
% \subfigure[$\ell=1.0 $ \AA]{\includegraphics[width=0.24\linewidth ]{HH2params_1p0.pdf}}
% \hfill
% \subfigure[$\ell=1.4 $ \AA]{\includegraphics[width=0.24\linewidth ]{HH2params_1p4.pdf}}
% \subfigure[$\ell=1.8$  \AA]{\includegraphics[width=0.24\linewidth ]{HH2params_1p8.pdf}}
% \subfigure[$\ell=2.2$ \AA]{\includegraphics[width=0.24\linewidth ]{HH2params_2p2.pdf}}
% \caption{H2 molecule with different bond lengths $\ell$. }
% \label{fig:H2}
% \end{figure*}

% \begin{figure*}[ht]
% \centering
% \subfigure[HeH bondlength=1.0]{\includegraphics[width=0.48\linewidth ]{HeH3params.pdf}}
% \hfill
% \subfigure[LiH 4 qubits]{\includegraphics[width=0.48\linewidth ]{LiH8params.pdf}}
% \caption{The estimated epectation energy as a function of the number of shots used for (a) the HeH and (b) the LiH molecules. }
% \label{fig:largesystem}
% \end{figure*}

\section*{Acknowledgement}
Y.C. and X.L. acknowledge the support to develop reduced scaling computational methods from the Scientific Discovery through Advanced Computing (SciDAC) program sponsored by the Offices of Advanced Scientific Computing Research (ASCR) and Basic Energy Sciences (BES) of the U.S. Department of Energy (\#DE-SC0022263).
This material is based upon work supported by the U.S. Department of Energy, Office of Science, Office of Advanced Scientific Computing Research and Office of Basic Energy Science, Scientific Discovery through Advanced Computing (SciDAC) program under Contract No. DE-AC02-05CH11231 and the Accelerated Research for Quantum Computing Program under Contract No. DE-AC02-05CH11231.  The authors thank Dr. Yuanran Zhu for helpful discussions on relevant topics.  This work used resources of the National Energy Research Scientific Computing Center (NERSC) using NERSC Award ASCR-ERCAP m1027 for 2023, which is supported by the Office of Science of the U.S. Department of Energy under Contract No. DE-AC02-05CH11231.

\bibliography{sample}

\end{document}